\documentclass[aps,prd,twocolumn,nofootinbib,showkeys,showpacs,floatfix]{revtex4-1}

\usepackage{amssymb,amsmath,amsfonts}
\usepackage{graphicx}
\usepackage{siunitx}
\usepackage{xcolor}
\usepackage[colorlinks=true, allcolors=blue]{hyperref}
\newcommand{\fb}{\femto\barn}
\DeclareSIUnit{\yr}{yr}

\newcommand{\AddrUFSM}{Universidad T\'ecnica Federico Santa Mar\'\i a,  \\ 
Casilla 110-V,  Valpara\'\i so,   Chile}
\newcommand{\AddrCCTVal}{Centro Cient\'\i fico Tecnol\'{o}gico de Valpara\'\i so,  \\ 
Casilla 110-V,  Valpara\'\i so,   Chile}

\begin{document}

\title{Left-right symmetric models at the high-intensity frontier}

\author{Oscar \surname{Castillo-Felisola}}
\author{Claudio O. \surname{Dib}}
\author{Juan C. \surname{Helo}}
\author{Sergey G. \surname{Kovalenko}}
\author{Sebastian E. \surname{Ortiz}}
\affiliation{\AddrUFSM,}
\affiliation{\AddrCCTVal.}


\begin{abstract}
We study constraints on left-right symmetric models from searches of semileptonic decays of 
$D$, $D_{s}$, and $B$ mesons, mediated by heavy neutrinos $N$ with masses $m_N\sim $ GeV that go on their mass shell leading to a resonant enhancement of the rates. Using these processes we examine, as a function of $m_N$ and $M_{W_R}$, the physics reach of the recently  proposed  high-intensity beam dump experiment SHiP, which is expected to  produce a large sample of $D_s$ mesons. We compare these results with the corresponding reach of neutrinoless double beta decay experiments, as well as like-sign dilepton searches  with displaced vertices at the LHC. We conclude that the SHiP experiment has clear advantages in probing the left-right symmetric models for heavy neutrinos in the \si{\GeV} mass range.
\end{abstract}

\maketitle

\section{Introduction}

One of the current avenues to search for new physics has been opened with the discovery of neutrino oscillations.\footnote{For the current status of oscillation data, see, for example, Ref.~\cite{Tortola:2012te}.} Oscillations indicate that neutrinos necessarily must be massive, while the standard model (SM) of electroweak interactions considers neutrinos as massless particles. Even if one tries to include the neutrino masses within the SM by means of Yukawa interactions ---the mechanism that gives mass to all other fermions--- the scenario does not come out natural, because one needs to introduce right-handed neutrino fields, which in turn lead to the possible inclusion of Majorana mass terms for these extra neutrino components, thus breaking the lepton number, and so on~\cite{Minkowski:1977sc,Mohapatra:1979ia,Schechter:1980gr}. In summary, the scenario opens into a wide range of possibilities, all of which indicate the existence of extra particles, mainly extra neutrinos, in a very broad range of energy scales~\cite{Atre:2009rg}.\footnote{The role of heavy neutrinos in astrophysics and cosmology has been studied in Refs.~\cite{Asaka:2005an,Asaka:2005pn,Bezrukov:2005mx,Boyarsky:2009ix,Drewes:2012ma,Sierra:2014sta,Lattanzi:2014mia,Falcone:2015hma}.} While each extension of the SM proposes extra neutrinos in a specific mass range, in general, the possibilities are from a few \si{\eV} all the way to grand unification scales of the order of \SI{e15}{\GeV}. On the other hand, the experiments can put bounds on specific combinations of neutrino masses and mixings, each one covering a different range of masses~\cite{Deppisch:2015qwa}.

In a typical seesaw model of neutrinos based on the SM gauge group, the heavy neutrinos are coupled to the standard sector through a small mixing with the standard leptons in the electroweak currents. In contrast, in a left-right symmetric gauge theory, the heavy neutrinos connect to the standard sector primarily through the right-handed currents. In the first case the couplings are suppressed by the mixing, while in the second case they are suppressed by the large mass of the $W_R$ bosons. In the present paper, we study the latter scenario, applied to rare decays of heavy mesons that are induced by massive neutrinos that go on mass shell in the intermediate state. Depending on the mass of the heavy neutrinos and the $W_R$ bosons, experiments could either discover the effect or establish new bounds, restricted to neutrino masses below the mass of a decaying particle, i.e., below \SI{5}{\GeV} for $B$ mesons or below \SI{1.8}{\GeV} for $D$ and $D_s$ mesons, so that the resonant effect can occur.

A particularly attractive feature of left-right symmetric (LR) models of electroweak interactions is that the appearance of right-handed neutrino components is not accidental but required in order to complete the right-handed lepton doublets~\cite{Pati:1974yy,Mohapatra:1974gc,Mohapatra:1980yp}. In these models, the heavy neutrinos are mainly the right-handed fields with small admixtures of the standard neutrinos. Consequently, they couple sizably to the SM sector through the gauge fields of the right-handed sector, $W_R$, but the latter induce very weak interactions on the SM particles due to the large scale of $M_{W_R}$.
The neutrino mass range of  interest here is appropriate for searches in $B$ factories such as BABAR~\cite{Lees:2013gdj} and BELLE~\cite{Liventsev:2013zz} and future high-intensity beam experiments such as SHiP~\cite{Bonivento:2013jag}. 

This article is organized as follows. In Sec.~\ref{TS} we briefly summarize the main formulas for production and decay of heavy neutrinos in LR models. In Sec.~\ref{sec:PLR} we discuss  the sensitivity of heavy neutrino searches driven by meson decays $D$, $D_s$, and $B$  in LR models as could be observed in the SHiP experiment, comparing them with current and future limits coming from neutrinoless double beta decay  ($0\nu \beta \beta$) experiments and equal-sign dilepton searches  with displaced vertices at the LHC. We leave the last section for a short discussion and summary.

\section{\label{TS}Theoretical Setup} 

Left-right symmetric models are well-motivated and  popular extensions of the standard model based on the gauge group  $SU(2)_L \times SU(2)_R \times U(1)_{\text{B-L}}$ with  gauge couplings $g_L$, $g_R$, and $g_1$, respectively, and a LR symmetric assignment of quarks and leptons: $Q_{L,R} = (u, d)_{L,R}$ and $L_{L,R} = (\nu, l)_{L,R}$~\cite{Pati:1974yy,Mohapatra:1974gc,Mohapatra:1980yp}. These theories have two particularly interesting features: First, LR symmetric models can be accommodated in broader groups such as Pati--Salam or $ SO (10) $; and second, these models contain three right-handed neutrinos, which are required to complete the right-handed lepton doublets, making it a  natural framework to justify the type-I seesaw mechanism. Thus, in a three-family scenario, there are three light (or standard)  and three heavy neutrino mass eigenstates, which here we call $N_i$. 

At some mass scale $M_{R}$, larger than the scale of the electroweak symmetry breaking, the gauge group $SU(2)_L \times SU(2)_R \times U(1)_{\text{B-L}}$ is broken down to the SM group $SU(2)_L \times U(1)_{\text{Y}}$. 
The charged current interactions relevant for the present analysis are
\begin{equation}
  \label{CC-NC-LR}
  \begin{split}
    \mathcal{L}
    &= \frac{g_{R}}{\sqrt{2}} \left( V^{(R)*}_{ud} \, \bar{d}  \gamma^{\mu} P_{R} u +  U^{(R)}_{l N}\cdot \bar l \gamma^{\mu} P_{R} N\  \right) W^-_{R \mu}  \\
    & +\frac{g_{L}}{\sqrt{2}} \left( V^{(L)*}_{ud}\bar{d}  \gamma^{\mu} P_{L} u +  U^{(L)}_{l N}\cdot \bar l \gamma^{\mu} P_{L} N \  \right) W^-_{L \mu}\, ,    
  \end{split}
\end{equation}
where $V^{(R)}$ and $V^{(L)}$ are the quark mixing matrices, and $U^{(R)}$ and $U^{(L)}$ are the lepton mixing matrices, of the left- and  right-handed sectors, respectively. In the rest of this work, we will assume $U^{(R)} = 1$ for simplicity, and thus we will simply denote $U^{(L)} = U$. The charged boson states $W_L^\pm$ and $W_R^\pm$ are linear combinations of mass eigenstates, denoted as $W_1^\pm$ and $W_2^\pm$:
\begin{equation}
  \label{W-MES}
  \begin{split}    
    W^{\pm}_{L} &= W^{\pm}_{1} \cos\zeta - W^{\pm}_{2} \sin\zeta,\\
    W^{\pm}_{R} &= W^{\pm}_{1} \sin\zeta + W^{\pm}_{2} \cos\zeta,
  \end{split}
\end{equation}
where the $W_L$-$W_R$ mixing angle $\zeta$, to a good approximation~\cite{MohapatraPal},  is given by
\begin{equation}
  \begin{split}    
    \tan 2\zeta &= \frac{2 g_{L}g_{R} M^{2}_{W_{L}} \sin 2\beta}{g_{R}^{2} M_{W_{L}}^{2} + g_{L}^{2}(M_{W_{R}}^{2} - M_{W_{L}}^{2})} \\
    &\approx  2  \frac{g_{R}}{g_{L}} \frac{M^{2}_{W_{L}}}{M^{2}_{W_{R}}} \sin 2\beta 
  \end{split}
  \label{WL-WR-mix}
\end{equation}
with $\tan\beta = \kappa^{\prime}/\kappa$ being the ratio of the two vacuum expectation values of the $SU(2)$ bidoublet Higgs $\Phi$.  Here  the approximate expression refers to $M_{W_{R}}\gg M_{W_{L}}$.  The masses of $W^{-}_{1,2}$ are denoted as $M_{W_{L}}$ and $M_{W_{R}}$, respectively. The maximal value of the $W_{L}$-$W_{R}$ mixing corresponds to $\kappa = \kappa^{\prime}$ when $\sin 2\beta = 1$.  The perturbativity condition $g_{R}^{2}/4\pi \leq 1$ together with $g_{L} = g(M_{Z}) \approx 0.64935$~\cite{Beringer:1900zz} sets the upper bound
\begin{equation}
  \label{WL-WR-mix-max}
  \tan 2\zeta \leq
  \frac{4 \sqrt{\pi}}{g(M_{Z})} \frac{M_{W_{L}}^{2}}{M_{W_{R}}^{2}}.
\end{equation}  

The total decay width of each of the three heavy neutrinos of the LR
models receives a negligible contribution from the neutral current 
interactions and can be written  in the mass range of  $m_N \sim$ \SIrange{1}{80}{\GeV}, to a good approximation, as~\cite{MohapatraPal,Maiezza:2010ic,Helo:2010cw}
\begin{equation}
  \label{Dec-Rate-RR}
  \Gamma_{N} 
  \approx \frac{3  G^{2}_{F}}{32 \pi^{3}} m_{N}^{5} 
  \left(\frac{M_{W_{L}}}{M_{W_{R}}}  \frac{g_{R}}{g_{L}}\right)^{4} 
  \left[1 + \sin^{2} 2\beta \right] %% \sum_{l} |U_{l N}|^{2}. 
\end{equation}
Here the masses of all final state particles are neglected.
For completeness we give the expression for the half-life $T_{1/2}$ of a neutrinoless double beta ($0\nu\beta\beta$) 
decay mediated by $W_R$ and heavy $N$ exchange~\cite{MohapatraPal,Maiezza:2010ic,Helo:2013esa}:
\begin{equation}
  \label{NLDBD-LR}
  T^{-1}_{1/2} =  G_{01} \left|{\mathcal{M}}_{N}\right|^{2} 
  \left| %\left(V_{eN}\right)^{2}
  \frac{m_{p}}{m_{N}}  
  \frac{M^{4}_{W_{L}}}{M^{4}_{W_{R}}} \frac{g^{4}_{R}}{g^{4}_{L}}
  \right|^{2},
\end{equation}
where $G_{01}$ is the phase space factor and ${\mathcal{M}}_{N}$ is the nuclear matrix element.  

\section{\label{sec:PLR}Phenomenology of LR symmetric models}

\begin{figure}[htb]
  \includegraphics[width=.9\linewidth]{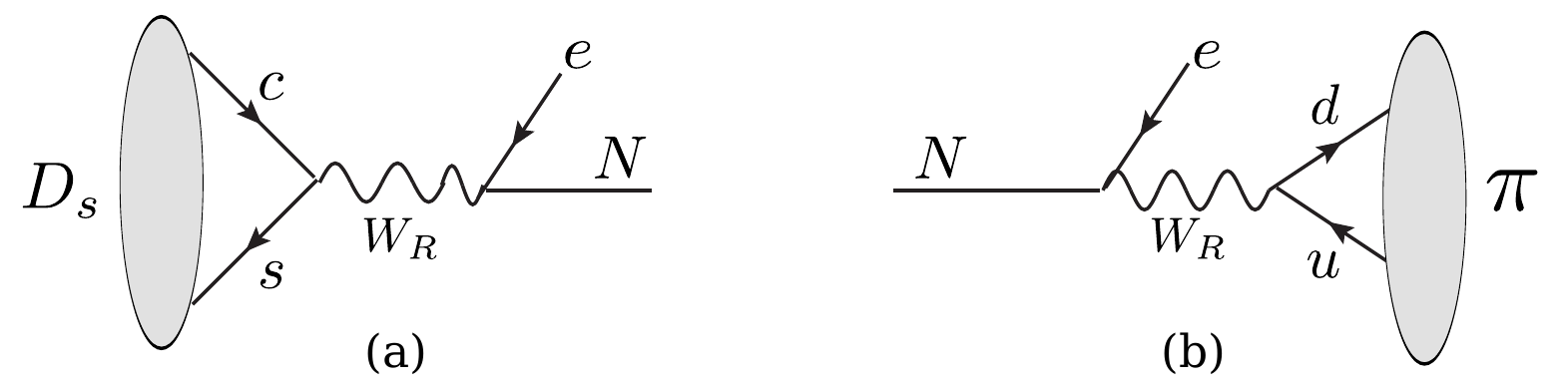}
  \caption{Heavy  neutrino production (a) and decay (b) in the LR symmetric model.}
  \label{fig:LR}  
\end{figure}

For simplicity, we will first restrict our discussion to heavy neutrino mixing with the electron sector only.  
We assume that the production  of a heavy Majorana neutrino in a meson decay, \mbox{$M \to e N$} (with $M$ either $D$, $D_s$, or $B$), and its subsequent decay as $N \to e\pi$, is
dominated by the $W_R$ boson exchange, as is shown in  Fig.~\ref{fig:LR}.

Consider a beam of heavy neutrinos $N$ with average energy $\bar E$, produced in a fixed target experiment. In the  manifestly LR symmetric scenario, i.e., \mbox{$g_R = g_L$,} the decay length of such neutrinos can be written according to Eq. (\ref{Dec-Rate-RR}) as a function of the  masses $m_N$ and $m_{W_R}$~\cite{Helo:2013esa} in the form
\begin{equation}
    \label{eq:decLR}
    L = c \bar{\gamma} \tau_{\!_N} \simeq  12 \, \bar{\gamma}  \, \left(\frac{\SI{1}{\GeV}}{m_N}\right)^5 \biggl(\frac{m_{W_R}}{\SI{1}{\TeV}}\biggr)^4 \;  [\si{\metre}],
\end{equation}
with $\bar{\gamma} = \bar{E} /m_N$. Therefore, for $m_{W_R}\sim $  \si{\TeV} and $m_N \sim \si{GeV}$ this decay length is rather large, and one can expect that only a fraction of the heavy neutrinos will decay inside a given detector.

In a LR symmetric model, the number of signal events in the form of meson decays $M \to e N$ followed by the heavy neutrino  decay $N \to e \pi $ are suppressed by a factor of $(m_{W_L}/m_{W_R})^8 $. This suppression comes from the production of  heavy neutrinos $M\to e N$, which  is suppressed by $(m_{W_L}/m_{W_R})^4$, and from the fraction of heavy neutrinos decaying inside the detector, which is proportional to $\tau_N^{-1}$, also suppressed by a factor of  $(m_{W_L}/m_{W_R})^4$. From the experimental point of view, this signal cannot distinguish a LR symmetric model  from a seesaw model based on the SM gauge group, where the number of signal events is suppressed by the small heavy-to-light neutrino mixing   $|U_{e N}|^2 \times |U_{e N}|^2 $~\cite{Bonivento:2013jag,  Alekhin:2015byh,Anelli:2015pba,  Dib:2014iga, Helo:2010cw} (see also~\cite{ Helo:2011yg, Dib:2011hc, Dib:2011jh, Cvetic:2015naa, Cvetic:2012hd, Cvetic:2010rw, Abada:2013aba, Quintero:2011yh, Dib:2000wm, Dib:2000ce}). 
However, for the same reason, if limits on $|U_{e N}|^2$  are known experimentally, we can extract   limits on   $m_{W_R}$,  by doing the conversion:
\begin{equation}
|U_{e N}|^2 \to \left( \frac{g_R}{g_L} \right)^4 \left( \frac{ m_{W_L} }{ m_{W_R} } \right)^4 \left| \frac{V^{(R)}_{cs}}{V^{(L)}_{cs}} \frac{V^{(R)}_{ud}}{V^{(L)}_{ud}} \right|,
\end{equation}
in the case of the  decay $D_s \to e N \to e e \pi$. For the \mbox{$B \to e N \to e e \pi$} one should replace $V_{cs}$ by $V_{ub}$ in the expression above. %%One can assume reasonable, that the
It is expected that the quark mixing matrices are similar $V^{(R)} \simeq V^{(L)}$~\cite{Senjanovic:2014pva,Senjanovic:2015yea}. Furthermore, in the manifestly LR scenario, i.e., $g_R = g_L$, %or  ... are equal ...
the conversion reduces to $|U_{e N}|^2 \to  ( { m_{W_L} }/{ m_{W_R} } )^4$.

\begin{figure}[tb]
  \begin{minipage}[b]{.95\linewidth}
    \includegraphics[width=.9\linewidth]{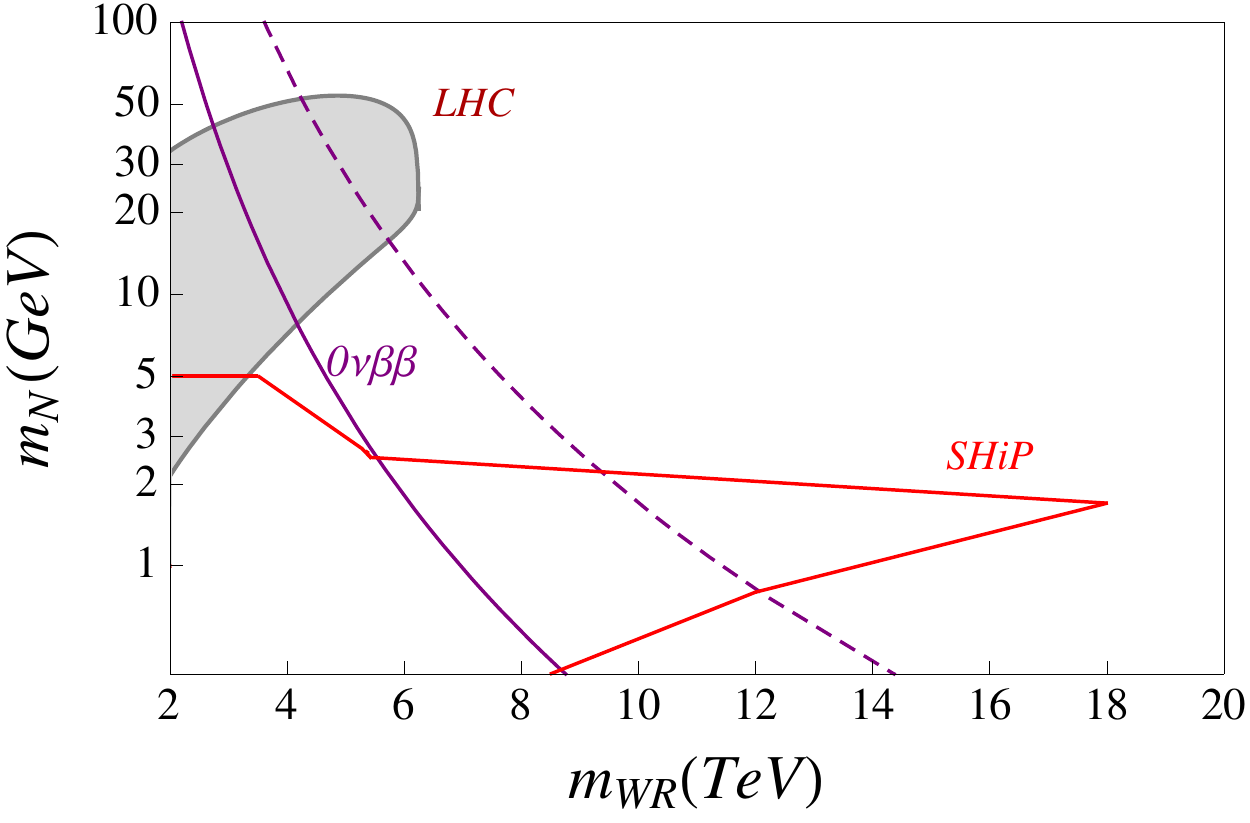}
    \includegraphics[width=.9\linewidth]{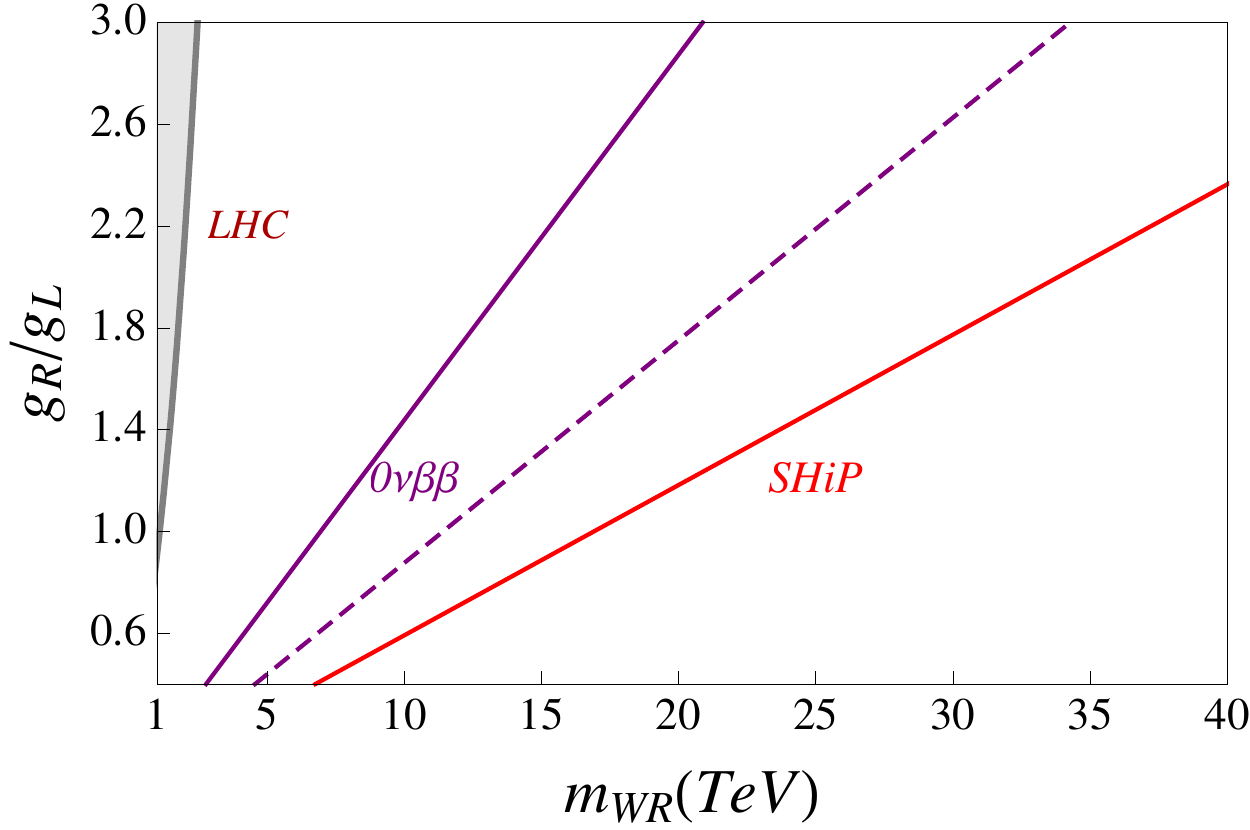} 
  \end{minipage}
  \caption{Regions in parameter space, which can be probed by a displaced vertex search  at the LHC,  $0 \nu \beta \beta$, and SHiP experiment. (a) Top: $m_{W_R}$ vs $m_N$ for fixed $g_R = g_L$. (b) Bottom: $g_R/g_L$ vs $m_{W_R}$ for fixed $m_N = \SI{1}{\GeV}$. For details, see the text.}
  \label{fig:LR-2}  
\end{figure}

The BELLE experiment, using $B$ meson decays $B \to X l N$ followed by $N \to e \pi$,   has set limits on $|U_{e N}|^2 \sim 10^{-4}$  in the heavy neutrino mass range \SI{0.5}-\SI{5}{\GeV}~\cite{Liventsev:2013zz}.\footnote{The LHCb experiment has searched for  heavy neutrinos by using the $B$ meson decay mode $B \to \mu N$ followed by $N \to \mu  \pi$  in a  mass range $m_N \approx$~\SIrange{0.5}{5}{\GeV}, setting limits on  $|U_{\mu N}|^2 \sim 10^{-3}$~\cite{Aaij:2014aba}.}  
In addition, the proposed  high-intensity beam  dump experiment SHiP~\cite{Bonivento:2013jag}, which  should produce a very  large number of $D_s$ mesons, 
will be sensitive to $|U_{e N}|^2 \sim 10^{-8}-10^{-10}$ for heavy neutrino masses of the order of $m_N \sim \SI{1}{\GeV}$~\cite{Deppisch:2015qwa,Alekhin:2015byh,Anelli:2015pba}.  In the same way,  these BELLE  searches on heavy neutrinos correspond to lower limits on  $m_{W_R}$  near 
$\SI{800}{\GeV}$, which are not better than the current limit  $m_{W_R} > \SI{2.5}{\TeV}$   coming from $K_0- \bar K_0$ mixing~\cite{Beall:1981ze,Zhang:2007fn,Zhang:2007da,Maiezza:2010ic,Senjanovic:2014pva,Senjanovic:2015yea}. On the other hand,  the future SHiP experiments will be sensitive to larger $W_R^\pm$ masses, up to $m_{W_R} \sim \SI{8}-\SI{18}{\TeV}$, for heavy neutrino masses $m_N \sim \SI{1}-\SI{2}{\GeV}$, as shown in Fig.~\ref{fig:LR-2}.   

\begin{figure}[tb]
  \begin{minipage}[b]{.95\linewidth}
  \includegraphics[width=.7\linewidth]{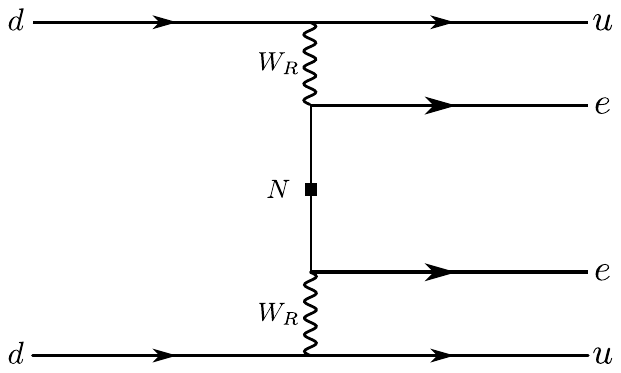}
  \vskip 1cm
    \includegraphics[width=.7\linewidth]{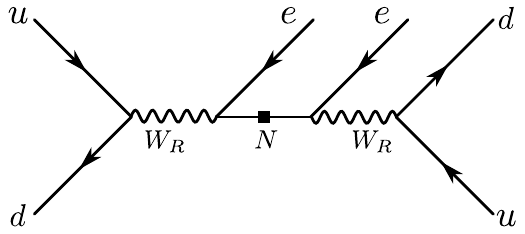}
      \end{minipage}
  \caption{ Top: Quark-level Feynman diagrams for left-right  realization of neutrinoless
double beta decay with heavy neutrino exchange. Bottom: The corresponding diagram at the LHC.}
  \label{fig:0nbb}
\end{figure}

The LHC can also constrain  LR symmetric models by searching  for  like-sign leptons plus two jets~\cite{Keung:1983uu,ATLAS:2012ak,Chen:2013foz,Khachatryan:2014dka}.\footnote{Generic processes of this kind have been discussed in Refs.~\cite{Helo:2012xe,Helo:2013ika, Helo:2013dla,Dev:2013wba} (see also~\cite{Bonnet:2012kh,Deppisch:2013jxa,Helo:2015fba}) as tree-level high-energy completions of lepton number violation operators that generate neutrinoless double beta decay.}  Currently, the LHC has imposed a lower limit for  $m_{W_R}$ at $\SI{3.0}{\TeV}$~\cite{Khachatryan:2014dka} in a neutrino mass range $m_N \sim$\SIrange{0.2}{2.0}{\TeV}. These limits   are  expected to be extended  up to $m_{W_R} \gtrsim  \SI{6}{\TeV}$~\cite{Senjanovic:2010nq} in future searches.  Although the  current searches at the LHC are not sensitive to heavy neutrinos with masses $m_N$ as low as a few \si{\GeV}, it is still possible  to search for heavy neutrinos  with masses $m_N \lesssim \SI{80}{\GeV}$ 
using displaced vertices~\cite{Helo:2013esa,Nemevsek:2011hz,Izaguirre:2015pga,Ng:2015hba,Humbert:2015yva}.\footnote{Recently, the authors of  Refs.~\cite{Maiezza:2015lza,BhupalDev:2012zg} have proposed to search for  heavy  neutrinos induced by Higgs decays at the LHC.} These heavy neutrinos could be produced  via a $W_R $ boson as is shown in Fig.~\ref{fig:0nbb}.   The advantage of a displaced vertex search is that, for a vertex separation roughly between $10^{-3}$ and $10^0$ m, there is little or no background \cite{Helo:2013esa}.  A signal based on this displaced vertex in principle could be affected by backgrounds from rare displaced hadron decays or from pileup effects \cite{Izaguirre:2015pga, TheATLAScollaboration:2013yia}. These backgrounds are found to be negligible for high track multiplicity and high enough $p_T$ cuts, which in our case do not eliminate the signal due to the large mass of the $W_R$.   

Using the narrow width approximation for the intermediate $N$ propagator, we can express the number of events of this type at the LHC decaying in the range from $d_1 = 1$~mm to $d_2=1$~m as
\begin{equation}
  \label{eq:events}
  \# (eejj) = \mathcal{L} \times \sigma(p p \to e N) \times Br(N \to e j j)
    \times P_N.
\end{equation}
Here $P_N = [ \exp(-d_1/L) - \exp(-d_2/L)]$ is  the probability for a  heavy neutrino to decay within a distance between $d_1 = 1$ mm and $d_2= 1$ m from its production point. Following the analysis  of  Ref.~\cite{Helo:2013esa}, we have used  Eq.~\eqref{eq:events} to  estimate the parametrical region that would produce at least five events with vertex separation  between $1$~mm and $ 1$~m, for an integrated luminosity of \mbox{$\mathcal{L} = \SI{300}{\per\fb}$} and collision energy $\sqrt{s}$\SI{= 13}{\TeV}, and  using the kinematical cuts in transverse momentum and rapidity $p_T \SI{> 40}{\GeV}$ and $\eta < 2.5$, respectively,  for both leptons and jets. 

In Fig.~\ref{fig:LR-2}(a), we show two limits from the nonobservation of $0 \nu \beta \beta$, assuming that the contribution via the $W_R$ boson and heavy neutrino exchange shown in Fig.~\ref{fig:0nbb} dominates. The weaker limit (solid line)~\cite{Maiezza:2010ic, Guadagnoli:2010sd, Hirsch:1996qw, Deppisch:2012nb} corresponds to the current bounds on the half-life of neutrinoless double beta decay, $T_{1/2} \SI{> 2e25}{\yr}$~\cite{Albert:2014awa, Agostini:2013mzu}, while the stronger limit  (dashed   line) corresponds to an expected future sensitivity of  $T_{1/2} \SI{>e27}{\yr}$. The grey band   bordered by  the solid  line shows the region  in parameter  space where a displaced vertex search  at the LHC could yield at least five events. The  solid line  near the bottom of Fig.~\ref{fig:LR-2}(a) shows  the region  in  the parameter space which can be probed by the heavy neutrino search at  SHiP. This is so far an unexplored domain of the $m_N-m_{W_R}$ parameter space below $m_N\sim 5$~GeV where, as seen from Fig.~\ref{fig:LR-2}(a), the SHiP searches are much more sensitive than the LHC and even more sensitive than optimistic future $0 \nu \beta \beta$ results for heavy neutrino masses $m_N \sim \SI{1}-\SI{2}{\GeV}$.

 Up to here, we have made the assumption that $g_R = g_L$, which corresponds  to  the   manifestly left-right symmetric scenario, a case that may not necessarily be so in general \cite{Arbelaez:2013nga}.  In Fig.~\ref{fig:LR-2}(b),  we show, just as in  Fig.~\ref{fig:LR-2}(a),  a comparison between  the $0 \nu \beta \beta$, LHC, and SHiP sensitivities, but for a fixed heavy neutrino mass $m_N = \SI{1}{\GeV}$ and in the plane  $g_R/g_L - m_{W_R}$. As shown, and in agreement with the previous analysis, the LHC is not competitive with  $0 \nu \beta \beta$  or SHiP for $m_N = \SI{1}{\GeV}$.  It is also clear that the SHiP experiment is capable to  probe a much larger area of this parameter space than even future  neutrinoless double beta decay  experiments.

 The lepton flavor violating muon    decays $\mu \to e \gamma$, $\mu \to e e e $, and $\mu \to e$ conversion in nuclei\footnote{The authors of Ref.~\cite{Domin:2004tk}  have studied the contribution of Majorana neutrinos in $\mu \to e$ conversion in nuclei.  See also Refs.~\cite{Faessler:2004jt,Faessler:2004ea,Gutsche:2009vp,Gutsche:2011bi}.} also  impose  constraints  on LR symmetric models.  However, for masses $m_N \sim  \si{\GeV}$ these bounds are very weak and not comparable to those  coming from  $0\nu \beta \beta$~\cite{ Das:2012ii,  Drewes:2015iva}.

Finally, we close by mentioning that so far we have considered heavy neutrino mixing only in the electron sector.    For   heavy neutrino mixing  in the $\mu$ sector, the LHC and SHiP limits shown in Fig.~\ref{fig:LR-2} will remain the same,  but  the limits coming from $0 \nu \beta \beta$ are not applicable. 
This is so because the  $0 \nu \beta \beta$ amplitude
is sensitive only to the $U_{eN}$ mixing matrix elements.
For   heavy neutrino mixing predominantly in the {\hbox{$\tau$-lepton}} sector, the situation is also different. On the one hand, the limits from the LHC will be much worse than the limits shown in Fig.~\ref{fig:LR-2}, due to the relatively poorer $\tau$ reconstruction efficiencies at the LHC.  On the other hand, the SHiP experiment   could be sensitive to $|U_{\tau N}|^2 \sim$~\numrange{e-8}{e-9} for heavy neutrino masses of the order of $m_N \sim \SI{1}{\GeV}$~\cite{Deppisch:2015qwa}, which correspond to strong  limits on   the $W_R$ boson mass  up to $m_{W_R} \sim \SI{9}{\TeV}$ for heavy neutrino masses $m_N \sim \SI{1}{\GeV}$ and $g_R = g_L$.

\section{\label{sec:SUM}Summary and Conclusions}

We have analyzed the scenario of heavy neutrinos within left-right symmetric models of electroweak interactions. We considered both a manifestly left-right symmetric case, as well as cases where the right and left gauge couplings are different. Concerning the heavy neutrino sector, we studied the cases where one neutrino is in the mass range $m_N$ from 1 to 80 GeV, and the rest are assumed to be heavier or not to interfere with the processes we consider. In LR symmetric models of neutrinos, the heavy neutrinos couple to the standard sector mainly through the right-handed currents, so their suppression is due to the large mass of the $W_R$ bosons. We studied the sensitivity of $0\nu\beta\beta$, LHC, and SHiP experiments to the mass of the heavy neutrinos and $W_R$ bosons. In the case of LHC, we considered signals of equal-sign dileptons with displaced vertices, which are free of backgrounds,
and in the case of SHiP, we considered the production of heavy neutrinos through charmed meson decays. We find that in the range of masses where SHiP is relevant, its sensitivity could be much stronger than the other two kinds of experiments, with the potential of discovery or considerable improvements on current $m_{W_R}$ and $m_N$ bounds. As a final comment, we should add that it could be possible to distinguish at SHIP a type-I seesaw scenario from a LR symmetric scenario by comparing the rates of semileptonic (i.e., $N\to e\pi$) with purely leptonic (i.e. $N\to e e \nu$) modes. Because of the strong suppression of $\nu$ production through the right-handed current, in the LR symmetric scenario the purely leptonic mode is highly suppressed, while in the type-I case the two modes are comparable. 

\begin{acknowledgments}
  We thank M. Hirsch  for  discussions about LR contributions to $0 \nu \beta \beta$ decay. This work was supported  by \mbox{Fondecyt (Chile)} Grants No. 11121557, No. 1150792,  and No. 1130617 and CONICYT (Chile) Grant No. 79140040.  
\end{acknowledgments}

\bibliographystyle{apsrev4-1}
\bibliography{Ref}

\end{document}